\documentclass{article}

\usepackage{arxiv}

\usepackage[utf8]{inputenc} 
\usepackage[T1]{fontenc}    
\usepackage{hyperref}       
\usepackage{url}            
\usepackage{booktabs}       
\usepackage{amsfonts}       
\usepackage{nicefrac}       
\usepackage{microtype}      
\usepackage{lipsum}
\usepackage{graphicx}
\usepackage{amsmath}
\usepackage{adjustbox}
\usepackage{multirow}
\usepackage{multicol}
\graphicspath{ {./images/} }

\title{Fast meningioma segmentation in T1-weighted MRI volumes using a lightweight 3D deep learning architecture}

\author{
 David Bouget \\
  Department of Medical Technology\\
  SINTEF\\
  Trondheim, Norway \\
  \texttt{david.bouget@sintef.no} \\
   \And
 Andr{\'e} Pedersen \\
  Department of Medical Technology\\
  SINTEF\\
  Trondheim, Norway \\
  \texttt{andre.pedersen@sintef.no} \\
  \And
 Sayied Abdol Mohieb Hosainey \\
  Department of Neurosurgery\\
  Bristol Royal Hospital for Children\\
  Bristol, United Kingdom \\
  \texttt{s.a.m.h@live.no} \\
    \And
 Johanna Vanel \\
  Department of Medical Technology\\
  SINTEF\\
  Trondheim, Norway \\
  \texttt{johanna.vanel@sintef.no} \\
     \And
 Ole Solheim \\
  Department of Neurosurgery\\
  St. Olavs hospital\\
  Trondheim, Norway \\
  \texttt{ole.solheim@ntnu.no} \\
     \And
 Ingerid Reinertsen \\
  Department of Medical Technology\\
  SINTEF\\
  Trondheim, Norway \\
  \texttt{ingerid.reinertsen@sintef.no} \\
}

\begin{document}
\maketitle
\begin{abstract}
Automatic and consistent meningioma segmentation in T1-weighted MRI volumes and corresponding volumetric assessment is of use for diagnosis, treatment planning, and tumor growth evaluation. In this paper, we optimized the segmentation and processing speed performances using a large number of both surgically treated meningiomas and untreated meningiomas followed at the outpatient clinic. We studied two different 3D neural network architectures: (i) a simple encoder-decoder similar to a 3D U-Net, and (ii) a lightweight multi-scale architecture (PLS-Net). In addition, we studied the impact of different training schemes. For the validation studies, we used 698 T1-weighted MR volumes from St. Olav University Hospital, Trondheim, Norway. The models were evaluated in terms of detection accuracy, segmentation accuracy and training/inference speed. While both architectures reached a similar Dice score of 70\% on average, the PLS-Net was more accurate with an F1-score of up to 88\%. The highest accuracy was achieved for the largest meningiomas. Speed-wise, the PLS-Net architecture tended to converge in about 50 hours while 130 hours were necessary for U-Net. Inference with PLS-Net takes less than a second on GPU and about 15 seconds on CPU. Overall, with the use of mixed precision training, it was possible to train competitive segmentation models in a relatively short amount of time using the lightweight PLS-Net architecture. In the future, the focus should be brought toward the segmentation of small meningiomas (less than 2ml) to improve clinical relevance for automatic and early diagnosis as well as speed of growth estimates.
\end{abstract}


\section{Introduction}
\label{intro}
Arising from the arachnoid caps cells on the outer surface of the meninges, meningiomas are the second most common primary brain tumor after gliomas, and account for approximately one-third of all central nervous system tumors~\cite{ostrom2019cbtrus}. With the increase in use of neuroimaging for checkups and precautionary diagnostics, incidental meningiomas are found more often~\cite{spasic2016incidental}.
Magnetic resonance imaging (MRI), adopted as the first routine examination, represents the gold standard for diagnosis and planning of the optimal treatment strategy (i.e.surgery or conservative management)~\cite{goldbrunner2016eano,kunimatsu2016variants}. While several different MR sequences may be used for meningioma imaging, measurements of tumor diameters and volumes are done using the contrast enhanced T1 weighted sequence.
Systematic and consistent segmentation of brain tumors is of utmost importance for accurate monitoring of growth and for guiding treatment decisions. With meningiomas being typically slow-growing tumors, performing detection at an early stage and monitoring systematically growth over time could improve clinical decision making and the patient's outcome~\cite{fountain2017volumetric}. Manual segmentation by radiologists, often in a slice-by-slice fashion is too time consuming to be part of daily clinical routine. Tumor volume and thus growth is therefore usually assessed based on manual measurements of tumor diameters resulting in considerable inter- and intra-rater variability~\cite{binaghi2016collection} and rough measures for growth evaluation~\cite{berntsen2020volumetric}. Automatic segmentation of pathology from MR images has been an active area of research for several decades but has made considerable progress with the recent advances in deep learning based methods~\cite{bauer2013survey,ueda2019technical}.
Nevertheless, the task of brain tumor segmentation remains challenging due to the large variability in appearance, shape, structure, and location~\cite{watts2014magnetic}. Similarly, problems might arise from the MRI volumes themselves whereby variability in resolution, intensity inhomogeneity~\cite{nyul2000new,tustison2010n4itk}, or varying intensity ranges among the same sequences and scanners can be noticed.
Gliomas, especially of low grade, are considered the most difficult brain tumors to segment in MRI since they often are diffuse, poorly contrasted, and with a tentacle-like structure. Conversely, typical meningiomas are sharply circumscribed with a strong contrast enhancement. However, smaller meningiomas may resemble other contrast enhancing structures, for example blood vessels (intensity, shape and size) particularly at the base of the brain, making them challenging to detect automatically.
In this study, we focus on the task of automatic meningioma segmentation using solely T1-weighted MRI volumes from both surgically treated patients and untreated patients followed at the outpatient clinic in order to create a method that is able to segment all tumor types and sizes.

\textbf{State-of-the-art:} As described in a recent review study~\cite{icsin2016review}, brain tumor segmentation methods can be classified into three categories based on the level of user interaction: manual, semi-automatic, and fully-automatic. For this study, we narrow the work to only fully-automatic methods specifically focused on  deep learning methods. In the past, a large majority of studies in brain tumor segmentation have been carried out using the Multimodal Brain Tumor Image Segmentation (BRATS) challenge dataset, which only contains glioma images~\cite{menze2014multimodal}. The task of brain tumor segmentation can be approached in 2D where each axial image (slice) from the original 3D MRI volume is processed sequentially. Havaei et al. proposed a two-pathway convolutional neural network (CNN) architecture to combine local and global information, arguing that the prediction for a given pixel should be influenced by both the immediate local neighborhood and a larger context such as the overall position in the brain~\cite{havaei2017brain}. Using the BRATS dataset for their experiments, they also proposed using a combination of all available MRI modalities as input for their method. In Zhao et al.~\cite{zhao2018deep}, the authors proposed to train CNNs and recurrent neural networks using image patches and slices along the three different acquisition planes (i.e. axial, coronal, sagittal) and fuse the predictions using a voting-based strategy. Both methods have been benchmarked on the BRATS dataset and were able to reach up to 80-85\% in terms of Dice coefficient and sensitivity/specificity. A large number of other studies have been carried out using image or image patch-based techniques as an attempt to deal with large MRI volumes in an efficient way~\cite{pereira2016brain,dvorak2015structured,zikic2014segmentation}. However, methods based on features obtained from image patches or across planes generally achieve lower performance than methods using features extracted from the entire 3D volume directly or through a slabbing process (i.e., using a set of slices). Simple 3D CNN architectures~\cite{myronenko20183d,isensee2018no}, multi-scale approaches~\cite{kamnitsas2017efficient,xu2018multi}, and ensembling of multiple CNNs~\cite{feng2020brain} have been explored. While they achieve better segmentation performances, are more robust to hyper-parameters and generalize better, the 3D nature of MRI volumes still poses challenges with respect to memory and computation limitations even on high-end GPUs.\\
While the availability of the BRATS dataset has triggered a large amount of work on glioma segmentation, meningioma segmentation has been less studied resulting in a scarce body of work. More traditional machine learning methods (e.g., SVM and graph cut) have been used for multi-modal (T1, T2) and multi-class (core tumor and edema) segmentation~\cite{binaghi2016collection}. While the reported performances are quite promising, the validation studies have been carried out on a dataset of only 15 patients. More recently, Laukamp et al. used different 3D deep CNN architectures (e.g., DeepMedic, BioMedIA) on their own multi-modal dataset~\cite{laukamp2019fully,laukamp2020automated}. While reported results reached above 90\% Dice score, the validation group consisted of only 56 patients. In addition, they investigated the use of heavy preprocessing techniques such as atlas registration and skull-stripping in combination with resampling and normalization. In their study, Pereira et al. also mentioned the effectiveness of normalization and data augmentation for brain tumor segmentation~\cite{pereira2016brain}.
A common limitation in the meningioma segmentation studies is the relatively limited number of patients included, and the choice of a fixed test set instead of a more thorough cross-validation approach. In general, the global trend in CNN architectures leads to ever larger and deeper 3D networks, even more so when considering ensembling strategies. As a consequence, the models' training and inference is becoming extremely computationally intensive, prohibiting their use in clinical settings with limited time and access only to regular computers.\\
In this paper, our contributions are: (i) the study of a lightweight 3D architecture that is less computationally intensive to use, (ii) a set of validation studies based on the largest meningioma dataset to date (698 patients), and (iii) an investigation into the trade-offs between segmentation performances and training/inference speed to enable clinical use.

\section{Data}
\label{sec:dataset}

\begin{figure}[ht]
\centering
\includegraphics[scale=0.85]{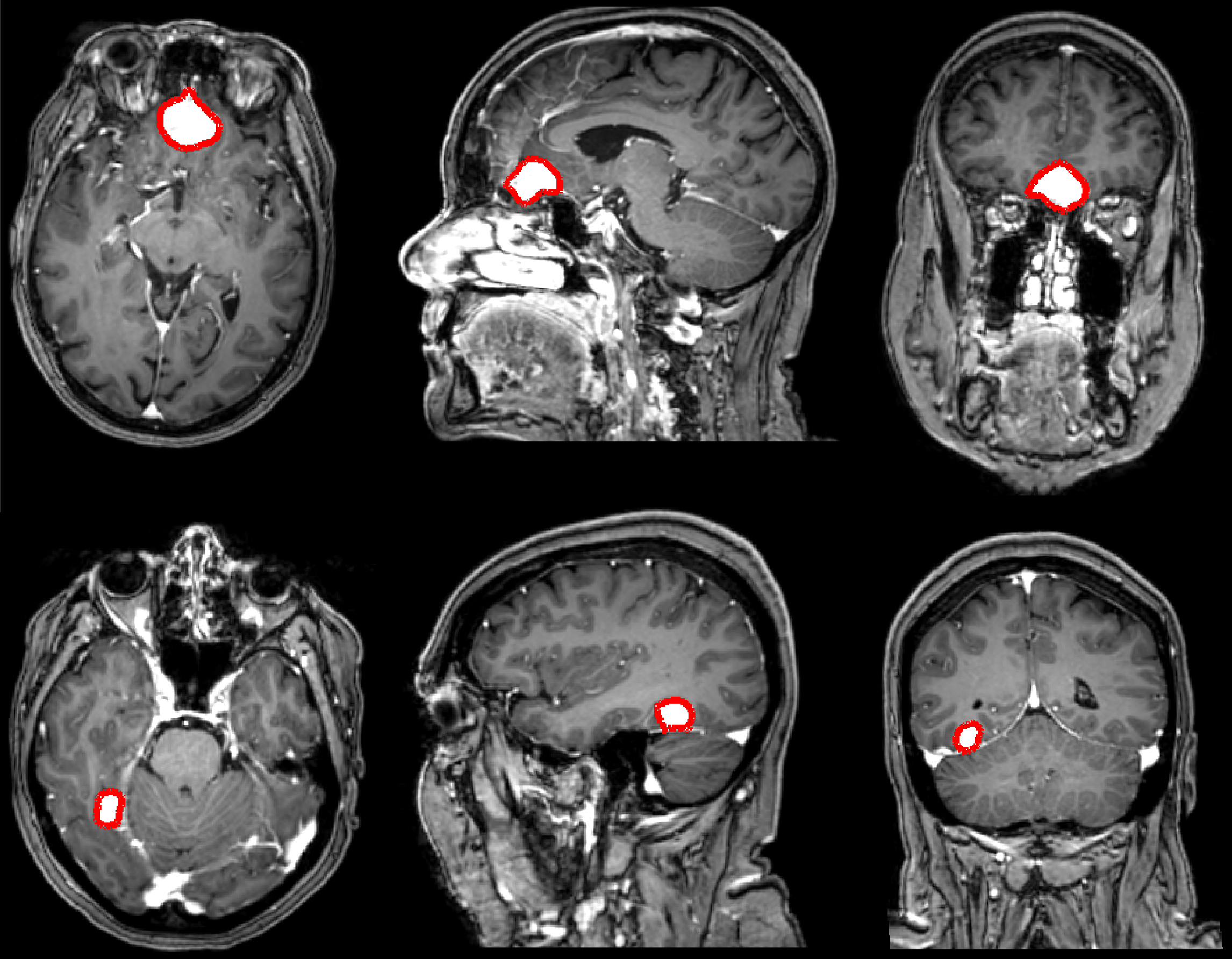}
\caption{Illustrations of the manually annotated meningiomas over the dataset (in red). Each row represents a different patient, and each column represents respectively the axial, coronal, and sagittal view.}
\label{fig:dataset-illu}
\end{figure}

For this study, we have used a dataset of 698 Gd-enhanced T1-weighted MRI volumes acquired on 1.5 or 3 Tesla scanners at one the seven hospitals in the geographical catchment region of the Department of Neurosurgery at St. Olavs University hospital, Trondheim, Norway between 2006 and 2015. All patients were 18 years or older with radiologically or histopathologically confirmed meningioma. Of those, 324 patients underwent surgery to remove the meningioma while the remaining 374 patients were followed at the outpatient clinic. Overall, MRI volume dimensions covered $[192; 512]\times[224; 512]\times[11; 290]$ voxels and the voxel sizes ranged between $[0.41; 1.05]\times[0.41; 1.05]\times[0.60; 7.00]\,\text{mm}^3$. All the meningiomas were manually delineated by an expert using 3D Slicer~\cite{pieper20043d}, and two examples are provided in Fig.~\ref{fig:dataset-illu}. Given the wide range in voxel sizes, especially in the z-dimension (slice thickness), we decided to further split our dataset in two. The first subset (DS1) consisted of the 600 high-quality MRIs with a slice thickness of at most 2.0\,mm, while the second subset (DS2) consisted of all 698 MRIs including the 98 images with a considerably higher slice thickness.
Overall, the meningiomas had a volume ranging $[0.07, 167.99]\,\text{ml}$.

We analyzed the differences between the groups of meningiomas. The volume of the surgically resected meningiomas was on average larger ($29.80\pm32.60\,\text{ml}$) compared to the untreated meningiomas followed at the outpatient clinic ($8.47\pm14.91\,\text{ml}$). A T-test showed statistical significance ($p<0.005$) between treatment strategy and tumor volume. Meningiomas for patients followed at the outpatient clinic are significantly smaller, making them more difficult to identify. Conversely, no statistical significance ($p=0.55$) has been unveiled between treatment strategy and poor image resolution. There were 50 MRIs with poor resolution for patients followed at the outpatient clinic and 48 MRIs for patient who underwent surgery.

\section{Methods}
\label{sec:methods}
First, we explain in Section~\ref{subsec:archs} our rationale for selecting the architectures and deep learning frameworks. Then we introduce in Section~\ref{subsec:preproc} the different preprocessing steps that can be applied. Finally, we present the selected training strategies for the two architectures in Section~\ref{subsec:train-strats}.

\subsection{Architectures and frameworks}
\label{subsec:archs}

\begin{figure}[h]
\centering
\includegraphics[scale=0.60]{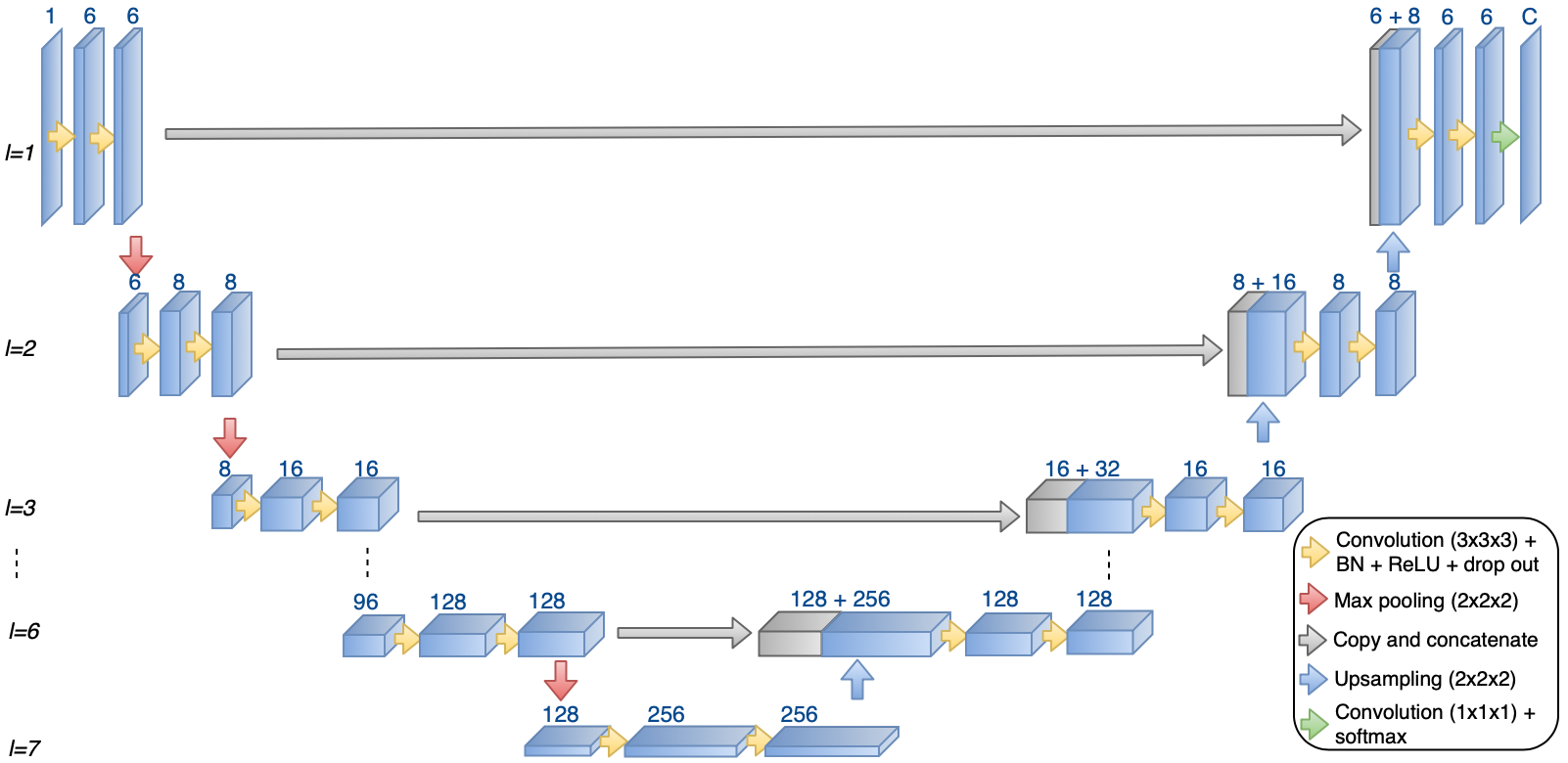}
\caption{3D U-Net architecture used in this study. The number of layers (l) and number of filters for each layer can vary based on input sample resolution.}
\label{fig:unet-arch}
\end{figure}

In early studies using fully convolutional neural network architectures, the original 3D MRI volumes were required to be split into 2D patches or slices before being processed independently and sequentially due to insufficient GPU memory. While it presented an advantage with respect to memory use, the lack of sufficient global information about the 3D relationships between voxels was detrimental for the overall performance. The advances in GPU design and increased memory capacity enabled the research on 3D neural network architectures to become mainstream.
For the task of semantic segmentation, encoder-decoder architectures have been favored, especially since the emergence of the U-Net~\cite{ronneberger2015u}, followed by the 3D U-Net~\cite{cciccek20163d}. Many U-Net variants have been studied in 2D and 3D for medical image segmentation over the past years and this architecture can be considered as a strong baseline~\cite{zhou2018unet++,alom2018recurrent,isensee2018nnu}. In this study, we have implemented an architecture close to the initial 3D U-Net, illustrated in Fig.~\ref{fig:unet-arch}. When working with 3D images, preprocessing is needed in order to fit the number of parameters on the GPU. Typical solutions are to either downsample the input volume, perform sub-division into slabs that are sequentially processed, or reduce the batch size to 1 which will result in poor  convergence. Training on mini-batches from size 2 to 32 have shown to improve generalization performances~\cite{masters2018revisiting}.

\begin{figure}[ht]
\centering
\includegraphics[scale=0.70]{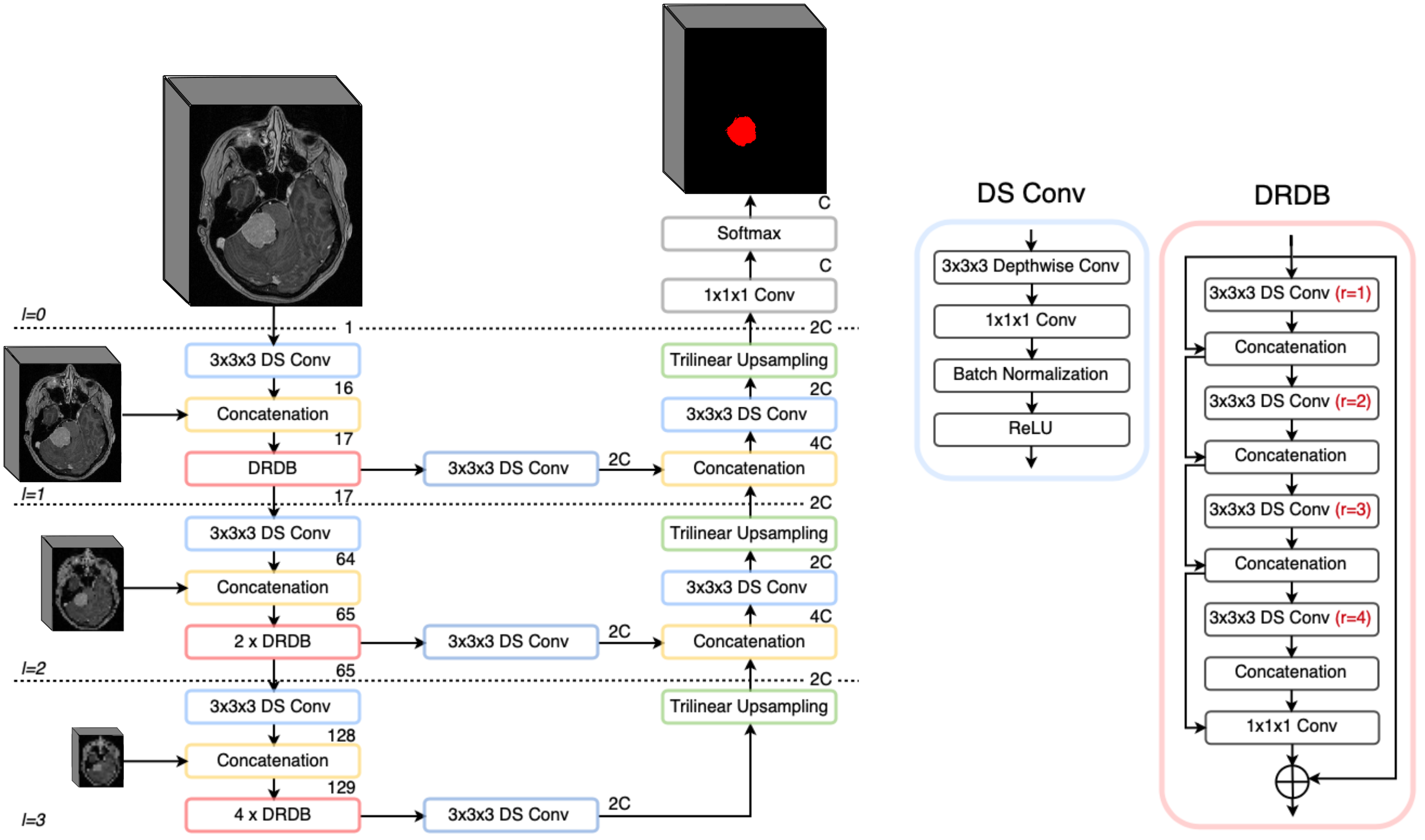}
\caption{PLS-Net architecture used in this study, kept identical as described in the original paper~\cite{lee2019efficient}.}
\label{fig:pls-arch}
\end{figure}

To take full advantage of the high-resolution MRI volumes as input, multi-scale encoder-decoder architectures have been proposed. Initially designed for the segmentation of lung lobes in CT volumes, the PLS-Net architecture is based on three insights: efficiency, multi-scale feature representation, and high-resolution 3D input/output~\cite{lee2019efficient}. The core components are (i) depthwise separable convolutions making the model lightweight and computationally efficient, (ii) dilated residual dense blocks to capture wide-range and multi-scale context features, and (iii) an input reinforcement scheme to maintain spatial information after downsampling layers. We have implemented the architecture as described in the original paper, and an illustration is provided in Fig.~\ref{fig:pls-arch}.

To address the issue of limited GPU memory, recent advances have been made for enabling the use of mixed precision computation rather than full precision for training neural networks. Mixed precision consists in using the full precision (i.e., float32) for some key specific layers (e.g., loss layer) while reducing most of the other layers to half precision (i.e., float16). The training process therefore requires less memory due to faster data transfer operations while at the same time math-intensive and memory-limited operations are sped up. These benefits are ensured at no accuracy expense compared to a full precision training. Since not all combinations of deep learning frameworks and GPU architectures are fully compatible with mixed precision training, we chose to use \texttt{TensorFlow}~\cite{abadi2016tensorflow} for full precision training, and \texttt{PyTorch}~\cite{paszke2019pytorch} for mixed precision training.

\subsection{Preprocessing}
\label{subsec:preproc}
In order to maximize and standardize the information input to the neural network, we propose a series of independent preprocessing steps for generating the training samples: 
\begin{itemize}
\item N4 bias correction using the \texttt{ANTs} implementation~\cite{avants2009advanced}.
\item Resampling to a uniform and isotropic spacing of 1\,mm using \texttt{NiBabel}, with a spline interpolation of order 1.
\item Cropping the volumes as tightly as possible around the patient's head by discarding the 20\% lowest intensity values (background noise) and identifying the largest remaining region. This is less restrictive and faster to perform than skull stripping.
\item Volume resizing to a specific shape dependent on the study/architecture using spline interpolation of order 1. When resizing based on an axial slice resolution, the new depth value is automatically inferred.
\item Finally, either normalization of the intensity to the range $[0, 1]$ (S) or zero-mean standardization (ZM).
\end{itemize}

\subsection{Training strategies}
\label{subsec:train-strats}
With the large range and inhomogeneous distribution of meningioma volumes, the baseline sampling strategy was to populate each fold with a similar volume distribution. We therefore split the meningiomas into three equally-populated bins and randomly sampled from these bins to generate the cross-validation folds.
All models were trained from scratch using the Adam optimizer with an initial learning rate of $10^{-3}$, the class-average Dice as loss function, and training was stopped after 30 consecutive epochs without validation loss improvement. All U-Net models were trained with a batch size of 8 using full precision in \texttt{TensorFlow} while all PLS-Net models were trained with batch size 4 using mixed precision with \texttt{PyTorch}.
We used a classical data augmentation approach where the following transforms have been applied to each input sample with a probability of 50\%: horizontal and vertical flipping, random rotation in the range $[-20^{\circ}, 20^{\circ}]$, translation up to 10\% of the axis dimension, zoom between $[80, 120]\%$, and perspective transform with a scale within $[0.0, 0.1]$. We selected two sets of augmentation methods: a minimalist approach with only flipping, rotation, and translation (Augm1); and an extended approach with all the above-mentioned transformations (Augm2).

\subsubsection{U-Net}

\begin{table}[h]
\centering
\caption{Overview of the different training strategies for the U-Net architecture.}
\begin{tabular}{r|r|r|r|r|r|r}
Configuration & Stride & Neg/Pos ratio & Norm. & Augm. & Resolution\tabularnewline
\hline
Cfg1 & 8 & None & S & Augm1 & $256\times192\times[167,420]$\tabularnewline
Cfg2 & 8 & 2.0 & S & Augm1 & $256\times192\times[167,420]$\tabularnewline
Cfg3 & 16 & 2.0 & S & Augm1 & $256\times192\times[167,420]$\tabularnewline
Cfg4 & 8 & 1.0 & S & Augm1 & $256\times192\times[167,420]$\tabularnewline
\end{tabular}
\label{tab:training-strats-unet}
\end{table}

As training strategy, we specifically investigated the impact of different sampling patterns in addition to the augmentation approach described above. Each patient's MRI volume was split into a collection of training samples (slabs) made of 32 slices along the z-axis. The stride parameter determined the number of slices shared by two consecutive slabs (i.e., an overlap of 24 slices for an stride of 8). Since some meningiomas are tiny, we also investigated balancing the ratio of positive to negative samples for each MRI volume. Random negative slabs were removed when the ratio was exceeded but no positive slab was excluded, purposely crafted as a non-bijective function. All MRI volumes were resized to an axial resolution of $256\times192$ pixels, leaving the third dimension adjusted dynamically following Eq.~\ref{eq:automatic-z}. For the architecture design, we used 7 layers with $[8, 16, 32, 64, 128, 256, 256]$ as number of filters and all spatial dropouts were set to a value of 0.1. In Table~\ref{tab:training-strats-unet}, we summarize the different configurations. 

\begin{equation}
    new\_dim_{z} = dim_{z} * \frac{new\_dim_{y}}{dim_{y}}
    \label{eq:automatic-z}
\end{equation}

\subsubsection{PLS-Net}
We decided to use the exact same architecture, number of layers, and kernel sizes as presented in the original paper. The single design choice was to keep a fixed input size of $256\times320\times224\, \text{pixels}$ while we focused on different preprocessing and data augmentation aspects, summarized in Table~\ref{tab:training-strats-pls}.

\begin{table}[h]
\centering
\caption{Overview of the different training strategies for the PLS-Net architecture.}
\begin{tabular}{r|r|r|r|r|r|r}
Configuration & Data & Bias Cor. & Norm. & Augm1 & Augm2 & Resolution\tabularnewline
\hline
Cfg1 & DS1 & False & S & True & False & $256\times320\times224$\tabularnewline
Cfg2 & DS1 & True & S & True & False & $256\times320\times224$\tabularnewline
Cfg3 & DS1 & False & ZM & True & False & $256\times320\times224$\tabularnewline
Cfg4 & DS1 & False & S & True & True & $256\times320\times224$\tabularnewline
Cfg5 & DS2 & False & S & True & True & $256\times320\times224$\tabularnewline
\end{tabular}
\label{tab:training-strats-pls}
\end{table}

\section{Validation studies}
\label{sec:validation}
In this work, we aim to optimize segmentation performances while finding the best trade-off with respect to processing speed. Unless specified otherwise, we followed a 5-fold cross-validation approach whereby at every iteration three folds were used for training, one for validation, and one for testing. 

\textit{Measurements:}
For quantifying the performances, we used: (i) the Dice score, (ii) the F1-score, and (iii) the training/inference speed.
The Dice score, reported in \%, is used to assess the quality of the pixel-wise segmentation by computing how well a detection overlaps with the corresponding manual ground truth.
The F1-score, reported in \%, assesses the combination of recall and precision performances.
Finally, the training speed (in $s.epoch^{-1}$), the inference speed IS (in ms), and the test speed TS (in $s.patient^{-1}$) to process one MRI are reported.

\textit{Metrics:}
For the segmentation task, the Dice score is computed between the ground truth and a binary representation of the probability map generated by a trained model. The binary representation is computed for ten different equally-spaced probability thresholds (PT) in the range $[0, 1]$. For the detection task, a similar range of probability thresholds is used to generate the binary results. A second threshold value (DT), in the list $[0, 0.25, 0.50, 0.75]$, is used to decide at the patient level if the meningioma has been sufficiently segmented to be considered a true positive, discarded otherwise (reported as Dice-TP). In case of multifocal meningiomas, a connected components approach coupled to a pairing strategy was employed to compute the recall and precision values.
Pooled estimates, computed from each fold's results, are reported for each measurement~\cite{killeen2005alternative}. Measurements are either reported with mean, mean and standard deviation, or mean and respective percentile confidence interval. If not stated otherwise, a significance level of 5 \% was used when calculating confidence intervals.

\textit{(i) Optimization study:} Performances using the different training configurations reported in Table~\ref{tab:training-strats-unet} and Table~\ref{tab:training-strats-pls} are studied. For U-Net, results are reported after training on the first fold only, given the time required to train one model.

\textit{(ii) Speed versus segmentation accuracy:}
This study aims at assessing which of the two architectures achieves the best overall performances considering all measurements, using the best configurations identified in the previous study.

\textit{(iii) Impact of dataset quality and variability:} Models trained with the best PLS-Net configuration were used for inference on the 98 low-resolution MRI volumes and the results were averaged. A direct comparison is done over the high-resolution and low-resolution images with models trained including the whole dataset (PLS-Cfg5).

\textit{(iv) Ground-truth quality:} In order to assess the quality of the manual annotations, all performed by a single expert, we performed an inter-annotator variability study. A random subset of 30 MRI volumes, 20 high-resolution and 10 low-resolution, was given for annotation to a second expert and differences were computed using the Dice score.

\section{Results}
\label{sec:results}

\subsection{Implementation details}
Results were obtained using an HP desktop: Intel Xeon @3.70 GHz, 62.5 GiB of RAM, NVIDIA Quadro P5000 (16GB), and a regular hard-drive. Implementation was done in Python using \texttt{TensorFlow} v1.13.1, and \texttt{PyTorch lightning} v0.7.3 with \texttt{PyTorch} back-end v1.3. For further training speed-up, all PLS-Net models were trained using the benchmark flag and Amp optimization level 2 (FP16 training with FP32 batch normalization and FP32 master weights). For data augmentation, all the methods used came from the Imgaug Python library~\cite{imgaug}.

\subsection{Optimization study}
\begin{figure}[!ht]
\centering
\includegraphics[scale=0.8]{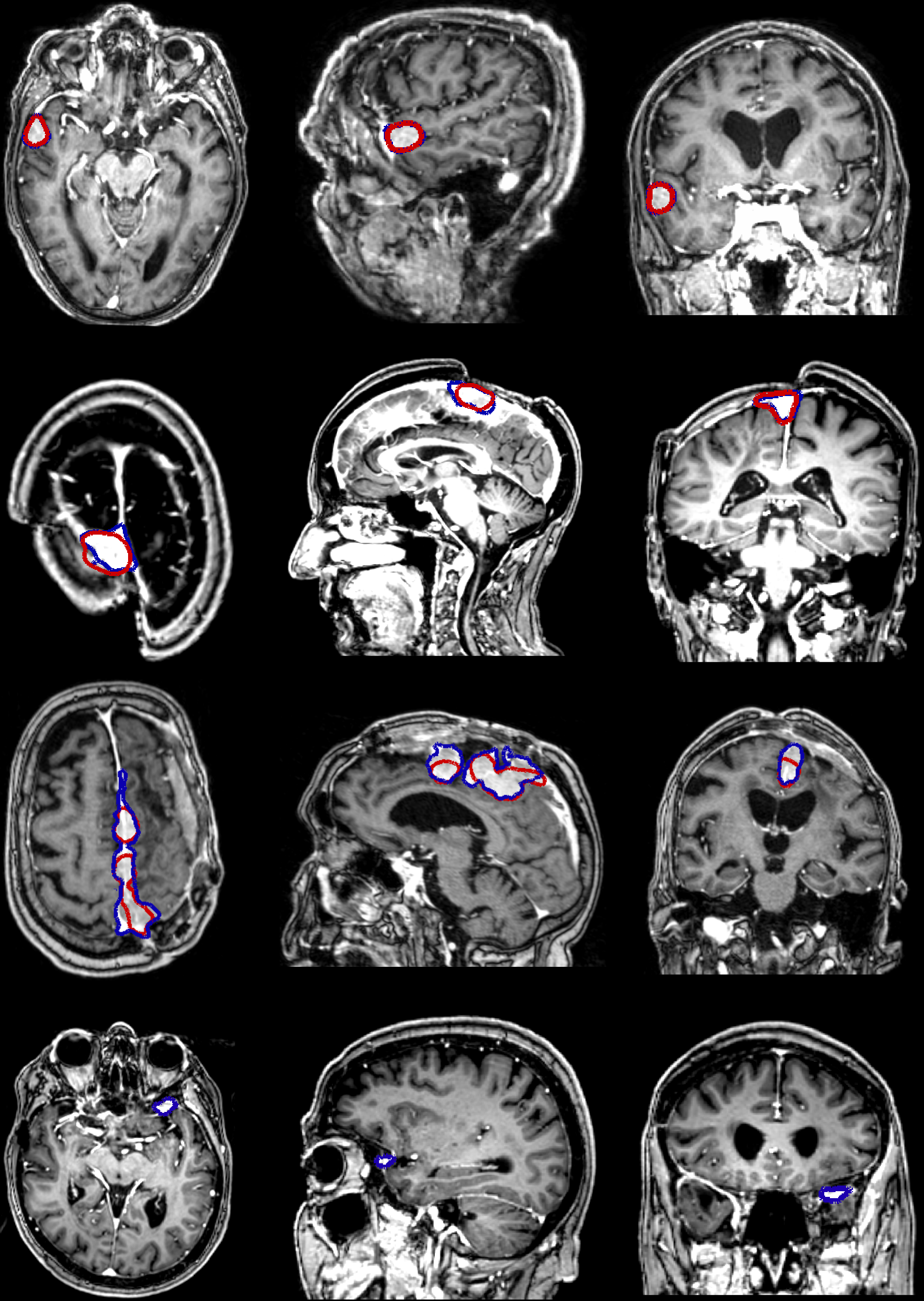}
\caption{Illustrations of segmentation results using the PLS-cfg4 model, each row representing a different patient. The ground truth for the meningioma in shown in blue whereas the automatic segmentation is shown in red.}
\label{fig:results-visual_examples}
\end{figure}
Results obtained for the U-Net configurations are reported in Table~\ref{tab:results-unet-opti}, while the ones for the PLS-Net architecture are reported in Table~\ref{tab:results-pls-opti}. With an optimized distribution of positive and negative training samples, U-Net performs similarly to PLS-Net regarding Dice performances. The highest precision is achieved with the first U-Net configuration, which was to be expected since all negative samples are kept. However, the best F1 score obtained with the U-Net architecture is far worse than with the PLS-Net architecture. The slabbing strategy generates more false positives since only local image features struggle to differentiate a small meningioma from other anatomical structures such as blood vessels. The different training configurations of PLS-Net provide comparable results across the board. The average Dice-TP reaches up to 87\%, indicating a good segmentation quality when a meningioma is detected. Considering the F1-score as the most important measurement for a relevant diagnosis use in clinical practice, UNet-cfg2 and PLS-cfg4 are the two best configurations.

\begin{table}[h]
\caption{Segmentation performances obtained with the different U-Net architecture configurations, over the first fold only.}
\centering
\adjustbox{max width=\textwidth}{
\begin{tabular}{c|c|c|c|c|c|c|c|c|c}
Cfg & PT & DT & Dice & Dice-TP & F1 & Recall & Precision \tabularnewline
\hline
Cfg1 & 0.6 & 0.5 & $63.49\pm36.65$ & $84.40\pm11.27$ & $77.13$ & $73.55$ & $81.06$\tabularnewline
Cfg2 & 0.6 & 0.25 & $67.75\pm33.99$ & $82.56\pm14.13$ & \boldmath{$77.78$} & $81.82$ & $77.76$\tabularnewline
Cfg3 & 0.4 & 0.25 & $69.27\pm32.76$ & $82.17\pm14.63$ & $76.27$ & $84.29$ & $69.63$ \tabularnewline
Cfg4 & 0.6 & 0.5 & \boldmath{$71.37\pm29.78$} & $84.19\pm10.13$ & $76.34$ & $81.82$ & $71.55$ \tabularnewline
\end{tabular}
}
\label{tab:results-unet-opti}
\end{table}

\begin{table}[h]
\caption{Segmentation performances obtained with the different PLS-Net architecture configurations, averaged across all folds.}
\adjustbox{max width=\textwidth}{
\begin{tabular}{c|c|c|c|c|c|c|c|c|c}
Cfg & PT & DT & Dice & Dice-TP & F1 & Recall & Precision \tabularnewline
\hline
Cfg1 & 0.5 & 0.5 & $73.40\pm31.34$ & $86.62\pm10.54$ & $88.01\pm1.39$ & $83.05\pm1.68$ & $93.68\pm2.47$ \tabularnewline
Cfg2 & 0.6 & 0.5 & $72.16\pm32.55$ & \boldmath{$87.19\pm9.40$} & $86.23\pm2.98$ & $80.74\pm2.43$ & $92.54\pm3.89$\tabularnewline
Cfg3 & 0.6 & 0.5 & \boldmath{$73.23\pm30.38$} & $86.01\pm10.47$ & $86.82\pm0.6$ & $82.90\pm2.1$ & $91.31\pm3.26$ \tabularnewline
Cfg4 & 0.5 & 0.25 & $71.69\pm33.41$ & $85.79\pm12.51$ & \boldmath{$88.34\pm1.86$} & \boldmath{$83.22\pm2.96$} & \boldmath{$94.19\pm1.06$} \tabularnewline
\end{tabular}
}
\label{tab:results-pls-opti}
\end{table}
Some examples, obtained with the PLS-cfg4 model, are displayed in Fig.~\ref{fig:results-visual_examples} where the ground truth is indicated in blue and the obtained segmentation is indicated in red. For the patients featured in the first two rows, the segmentation is almost perfect, while for the third patient the whole extent of the meningioma is not fully segmented. In the last case, the meningioma is both relatively small and located right behind the eye socket, and as such has not been detected at all.

\begin{figure}[ht]
\centering
\includegraphics[scale=0.50]{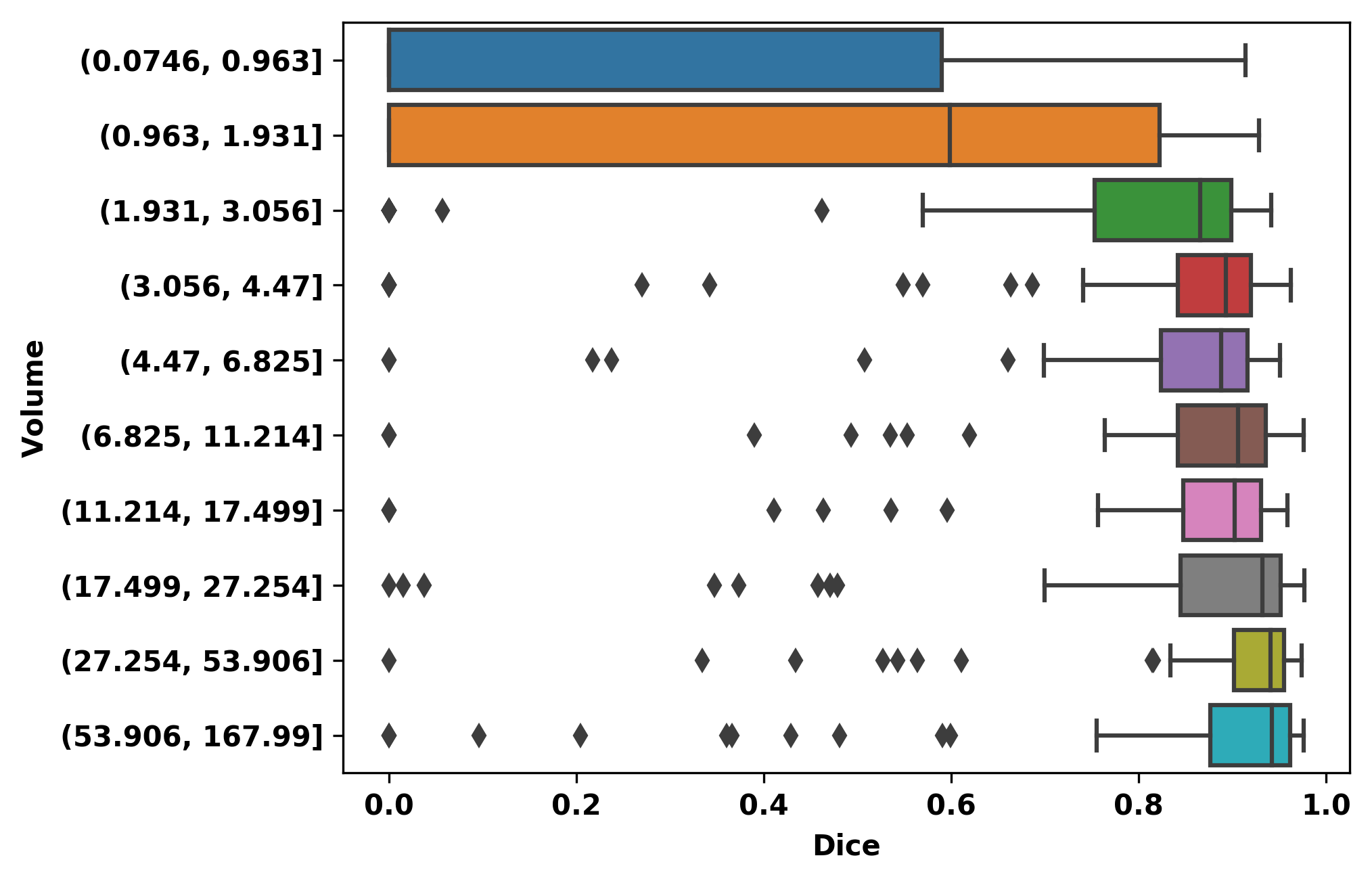}~
\includegraphics[scale=0.47]{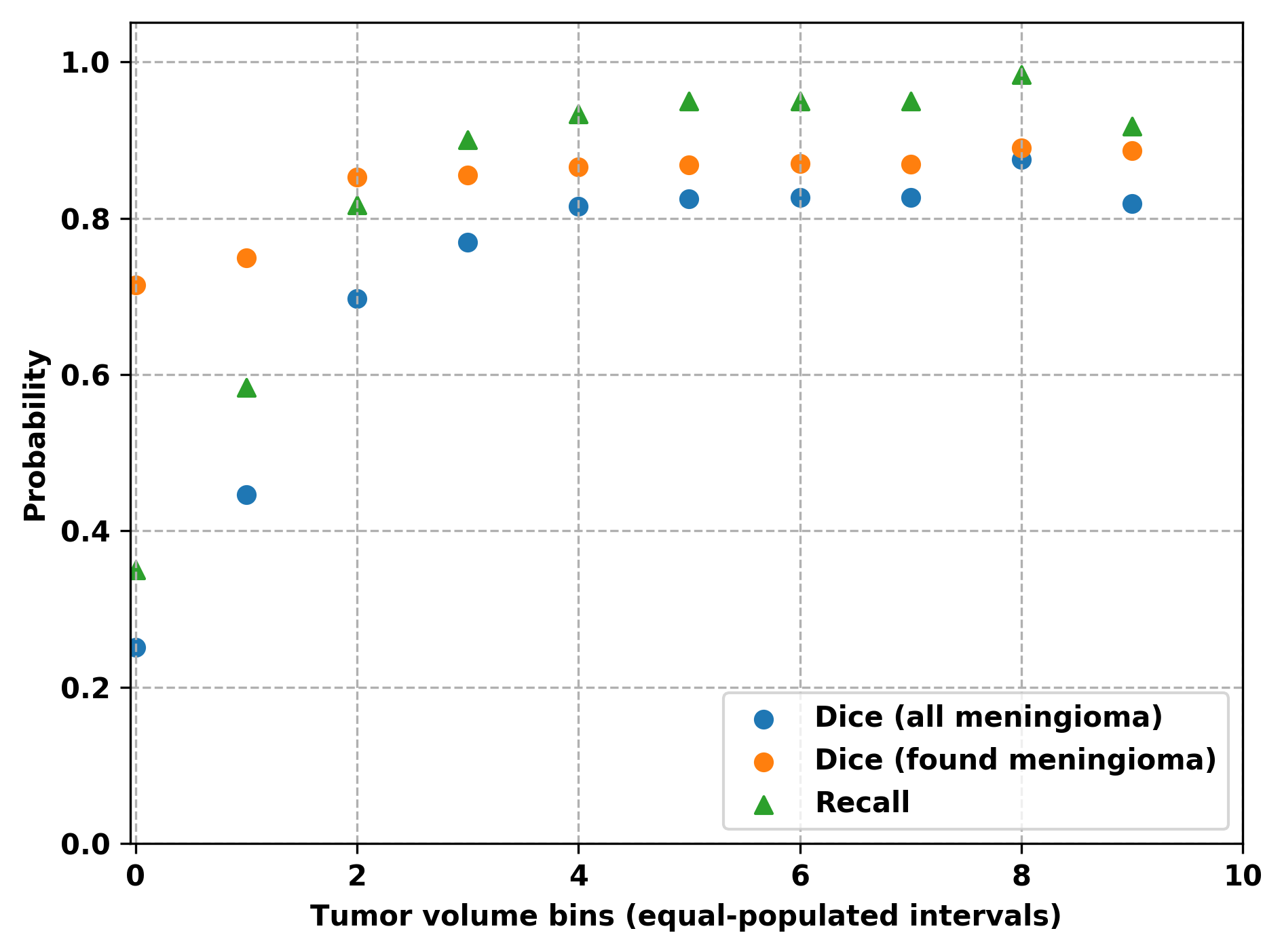}\\\includegraphics[scale=0.50]{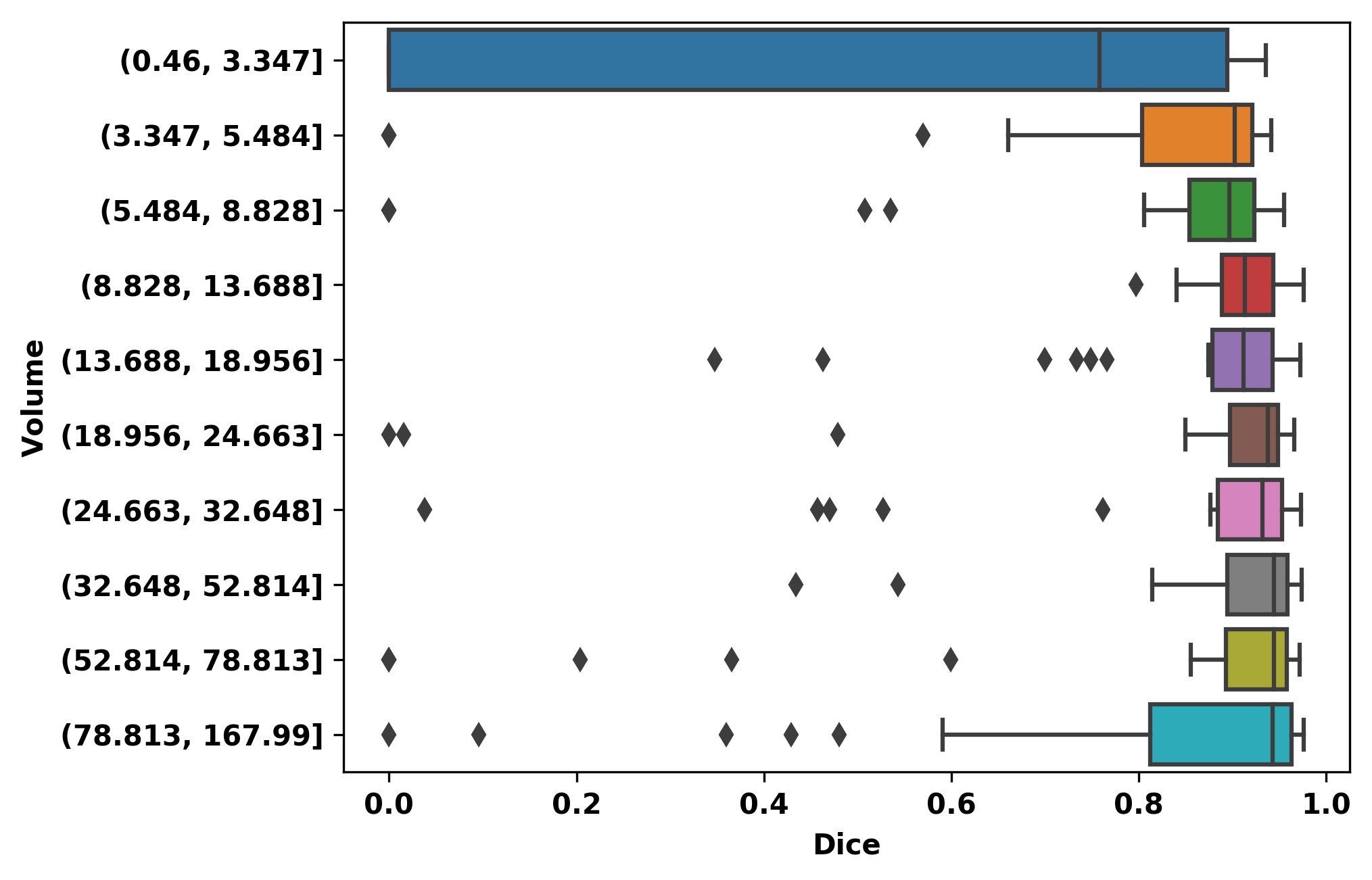}~
\includegraphics[scale=0.47]{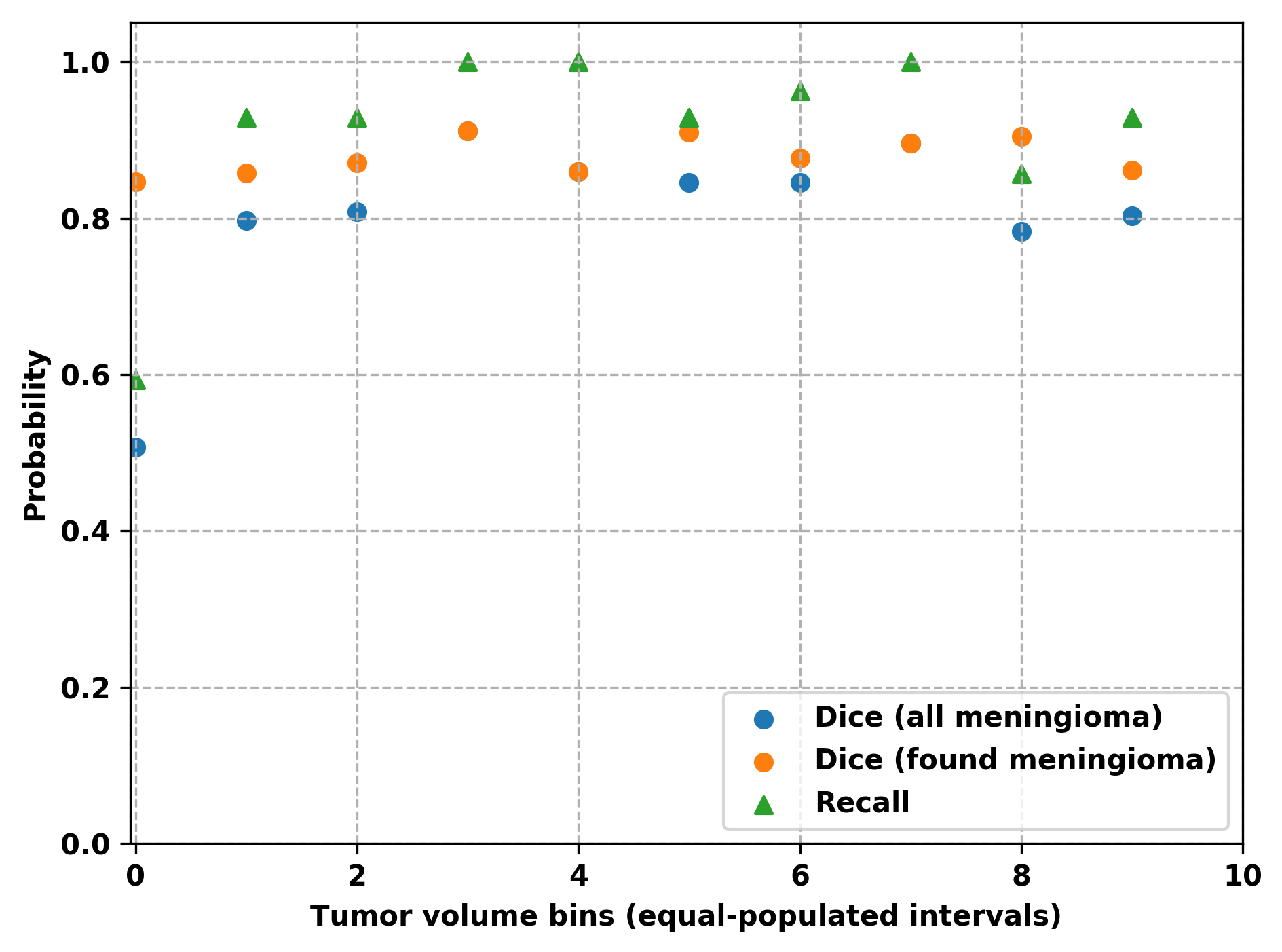}\\
\includegraphics[scale=0.50]{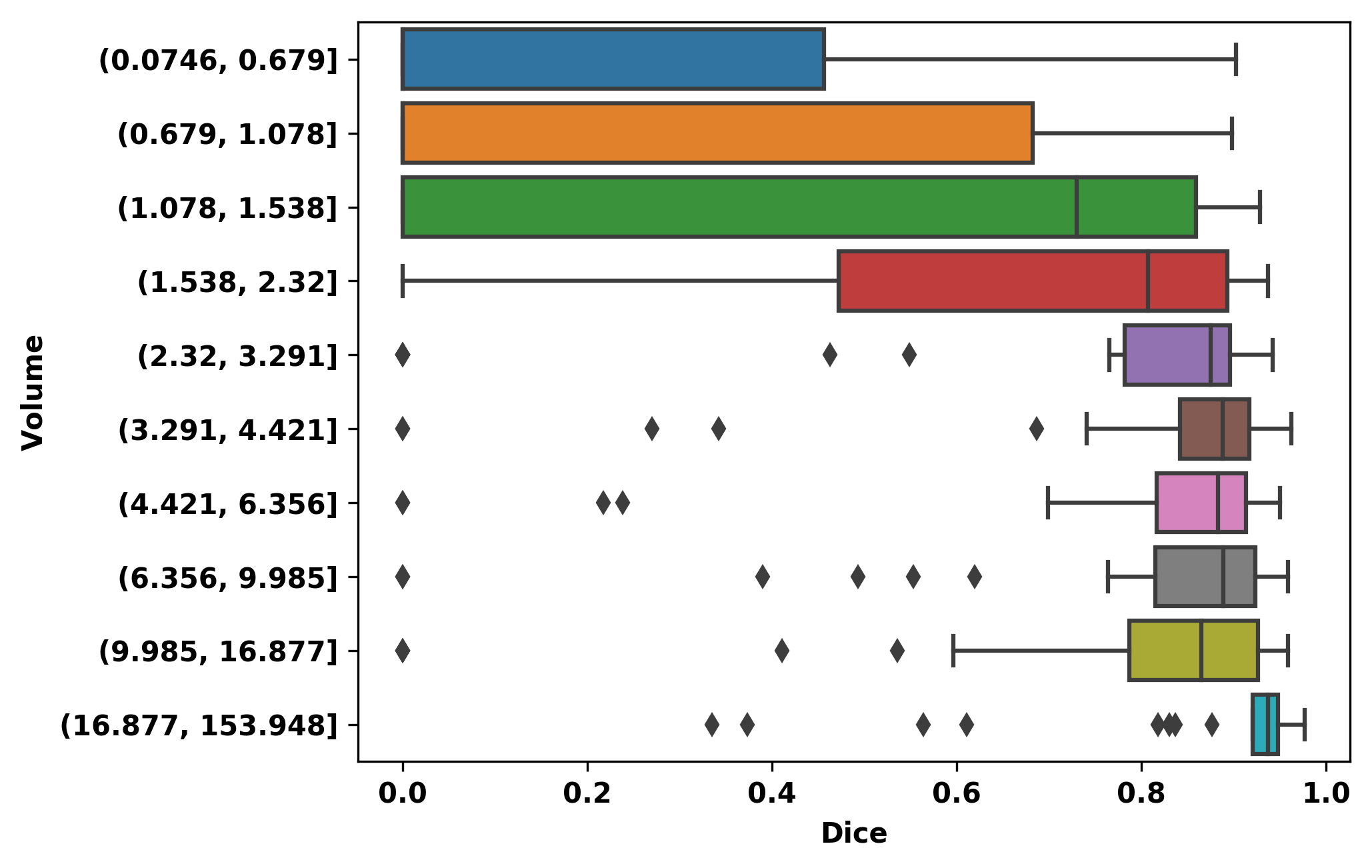}~\includegraphics[scale=0.47]{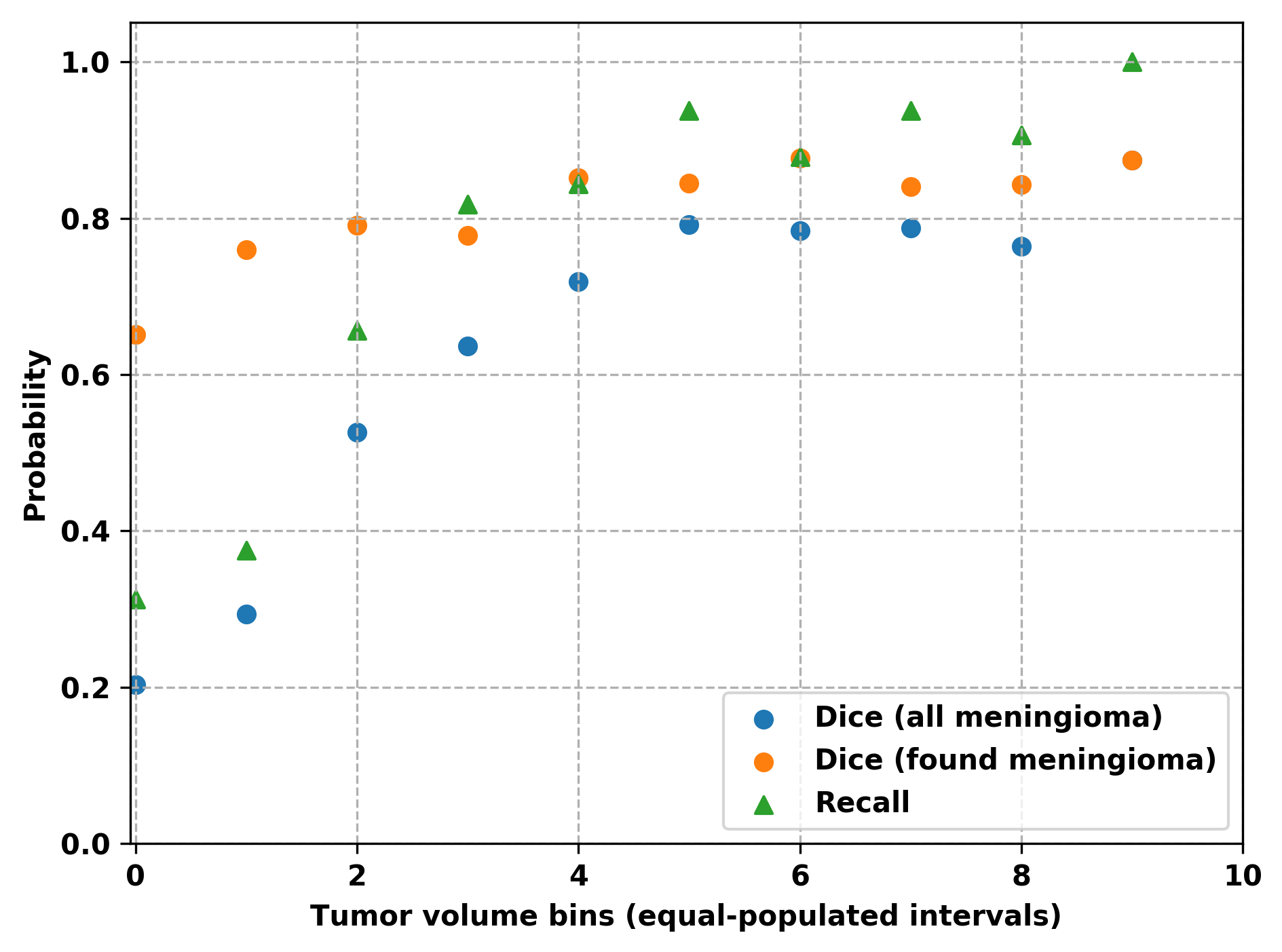}
\caption{Overall (top), hospital (middle), and outpatient clinic (bottom) results for PLS-cfg4. The first column shows Dice performances over tumor volumes, the second column shows Dice, Dice-TP and recall performances over tumor volumes. Ten equally populated bins, based on tumor volumes, have been used to group the meningioma performances.}
\label{fig:results-pls-volumeprov}
\end{figure}

When considering the origin of the data (i.e., hospital or outpatient clinic) reported in Table~\ref{tab:results-pls-origin}, performance appears to be better for surgically treated tumors reaching an F1-score higher than 90\% with the PLS-Net architecture. Conversely, more meningiomas from the outpatient clinic are left undetected and thus unsegmented, explaining the lower recall and average Dice score. Meningiomas from outpatient clinic patients being statistically smaller than surgically treated meningiomas, we further analyzed the relationship between the treatment strategy and tumor volume as shown in Fig~\ref{fig:results-pls-volumeprov}. Small meningiomas ($<2\,\text{ml}$) are challenging to segment and are either not or poorly segmented with at best a 50\% Dice score. For larger meningiomas ($>3\,\text{ml}$) the Dice score reaches 90\% whether surgically resected or followed at the outpatient clinic. Segmentation performance is heavily impacted by meningioma volumes, and probably also by the larger variability in tumor location within the brain for those smaller meningiomas. This is a clear indication that designing better sampling strategies for training is of utmost importance to train a robust and generic model.

\begin{table}[ht]
\caption{Best segmentation performances based on the treatment strategy: surgery or follow-up at the outpatient clinic.}
\adjustbox{max width=\textwidth}{
\begin{tabular}{c|c|c|c|c|c|c|}
Cfg & Origin & Dice & Dice-TP & Recall & Precision & F1 \tabularnewline
\hline
\multirow{2}{*}{UNet-cfg2} & Hospital & $82.47\pm22.57$ & $88.66\pm8.11$ & $91.42\pm3.91$ & $78.44\pm3.86$ & $84.36\pm2.95$ \tabularnewline
& Clinic & $64.85\pm32.16$ & $81.79\pm10.49$ & $75.06\pm9.45$ & $76.79\pm5.36$ & $75.79\pm7.17$ \tabularnewline
\hline
\multirow{2}{*}{PLS-cfg4} & Hospital & $81.29\pm26.08$ & $88.81\pm9.82$ & $91.41\pm1.22$ & $93.61\pm2.22$ & $92.49\pm1.58$\tabularnewline
& Clinic & $63.44\pm36.50$ & $82.53\pm14.16$ & $75.82\pm6.51$ & $95.05\pm1.12$ & $84.17\pm3.73$ \tabularnewline
\end{tabular}
}
\label{tab:results-pls-origin}
\end{table}

\subsection{Speed versus segmentation accuracy}
On average, convergence is achieved much faster with the PLS-Net architecture ($<50\,hours$) than with U-Net ($130\,hours$) when leaving enough room for the models to grind. Competitive models with a validation loss below 0.2 can even be generated in shorter time using the PLS-Net architecture ($<20\,hours$). A summary of the training time and convergence speed is presented in Table~\ref{tab:results-training-speed}.
The inference speed is fast with both architectures, making them both usable in practice. On average with the U-Net architecture, the inference speed is of $3.58\pm0.22\texttt{s}$ for a total processing time of $21.48\pm7.89\texttt{s}$. With this architecture, the MRI volume is split into non-overlapping slabs that are processed sequentially. For the PLS-Net architecture, the inference speed is lowered to $950\pm14\texttt{ms}$ for a total processing time of $14.15\pm4.5\texttt{s}$. The small number of trainable parameters with PLS-Net (0.251\,M) also makes it usable on low-end computers simply equipped with a CPU. In comparison, our U-Net architecture is made of 14.75\,M trainable parameters, which is consequently higher. In case of pure CPU usage, the total processing time of a new MRI with the PLS-Net architecture increases to $135\pm10\texttt{s}$.

\begin{table}[h]
\caption{Training time results for the different U-Net and PLS-Net configurations. Results are averaged across the five folds when possible.}
\centering
\adjustbox{max width=\textwidth}{
\begin{tabular}{r|r|r|r|r|}
Cfg & \# samples & s.$epoch^{-1}$ & Best epoch & Train time (hours)\tabularnewline
\hline
UNet-Cfg1 & $19\,684$ & $5\,800$ & $79$ & $127.28$\tabularnewline
UNet-Cfg2 & $14\,617$ & $3\,990$ & $120\pm40$ & $132.78\pm44.18$ \tabularnewline
UNet-Cfg3 & $7\,359$ & $2\,120$ & $153$ & $90.1$ \tabularnewline
UNet-Cfg4 & $10\,321$ & $2\,860$ & $105$ & $83.42$\tabularnewline
PLS-Cfg1 & $600$ & $1\,920$ & $113\pm18$ & $60.27\pm9.56$ \tabularnewline
PLS-Cfg2 & $600$ & $1\,920$ & $106\pm27$ & $56.74\pm14.23$ \tabularnewline
PLS-Cfg3 & $600$ & $1\,920$ & $86\pm29$ & $45.97\pm15.47$ \tabularnewline
PLS-Cfg4 & $600$ & $1\,920$ & $91\pm23$ & $48.75\pm12.48$ \tabularnewline
PLS-Cfg5 & $698$ & $2\,220$ & $91\pm31$ & $56.00\pm19.04$
\end{tabular}
}
\label{tab:results-training-speed}
\end{table}

\subsection{Impact of input resolution}
\begin{figure}[ht]
\centering
\includegraphics[scale=0.8]{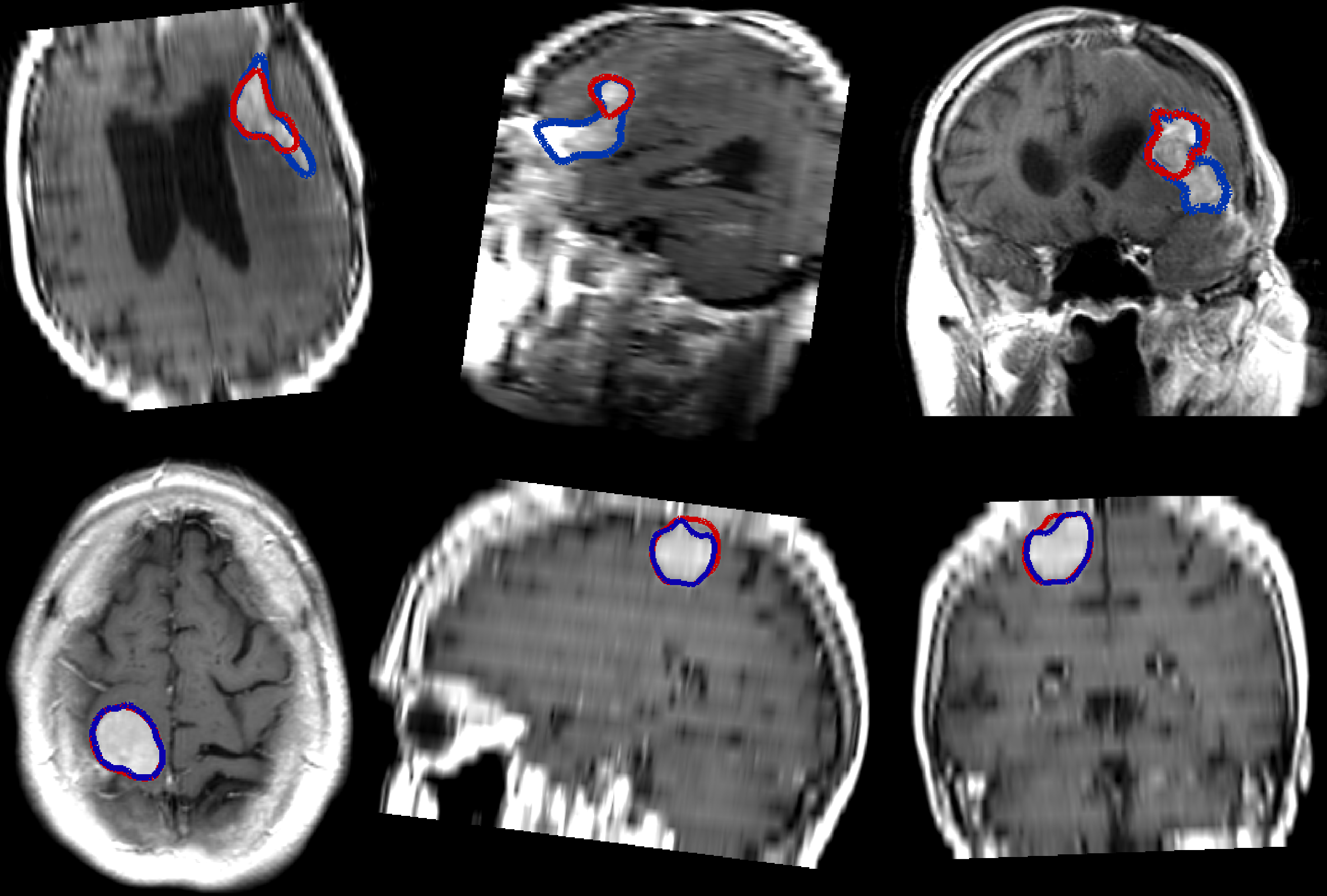}
\caption{Illustrations of segmentation results on images with a poor resolution using the PLS-cfg4 model, each row representing a different patient. The ground truth for the meningioma in shown in blue whereas the automatic segmentation is shown in red.}
\label{fig:results_poorres-visual_examples}
\end{figure}

Segmentation performances for the high- and low- resolution images are summarized in Table~\ref{tab:results-pls-slice-thickness}. Only minor differences across all performance metrics can be seen whether the low-resolution images are used during the training process or left aside. Selecting only the high-resolution images for training, coupled to advanced data augmentation methods, allows the trained models to be robust to extreme image stretching when resizing an MRI with for example an original slice thickness of 5\,mm. Figure~\ref{fig:results_poorres-visual_examples} provides segmentation results on low-resolution images.

\begin{table}[ht]
\caption{Performances analysis when including the images with a slice thickness larger than 2\,mm.}
\adjustbox{max width=\textwidth}{
\begin{tabular}{c|c|c|c|c|c|c|}
Resolution & Cfg & Dice & Dice-TP & Recall & Precision & F1 \tabularnewline
\hline
\multirow{2}{*}{High} & PLS-cfg4 &  $71.69\pm33.41$ & $85.79\pm12.51$ & $83.22\pm2.96$ & $94.19\pm1.06$ & $88.34\pm1.86$\tabularnewline
& PLS-cfg5 & $73.19\pm32.31$ & $87.28\pm9.44$ & $82.65\pm4.25$ & $95.25\pm1.63$ & $88.43\pm2.36$ \tabularnewline
\hline
\multirow{2}{*}{Low} & PLS-cfg4 & $61.09\pm33.15$ & $78.88\pm12.71$ & $74.69\pm5.59$ & $85.39\pm3.04$ & $79.61\pm3.72$ \tabularnewline
& PLS-cfg5 & $62.70\pm33.34$ & $81.34\pm12.55$ & $73.70\pm8.65$ & $86.57\pm6.89$ & $79.39\pm6.81$ \tabularnewline
\end{tabular}
}
\label{tab:results-pls-slice-thickness}
\end{table}

\subsection{Ground truth quality}
Between the two experts, the segmentation is matching with an average Dice score of 89.1 [86.3, 92.0], indicating a strong similarity. The Dice was higher for the high-resolution scans, with a Dice of 92.0 [89.8, 94.2], compared to 83.4 [75.8, 91.0] for the low-resolution ones. However, as the confidence intervals overlap, there is not a significant difference between the annotators with respect to image resolution. There was also found no difference in the models performances on both ground truths. This indicates that the initial ground truth is sufficient for training good models in terms of segmentation.

\begin{figure}[!ht]
\centering
\includegraphics[scale=0.6]{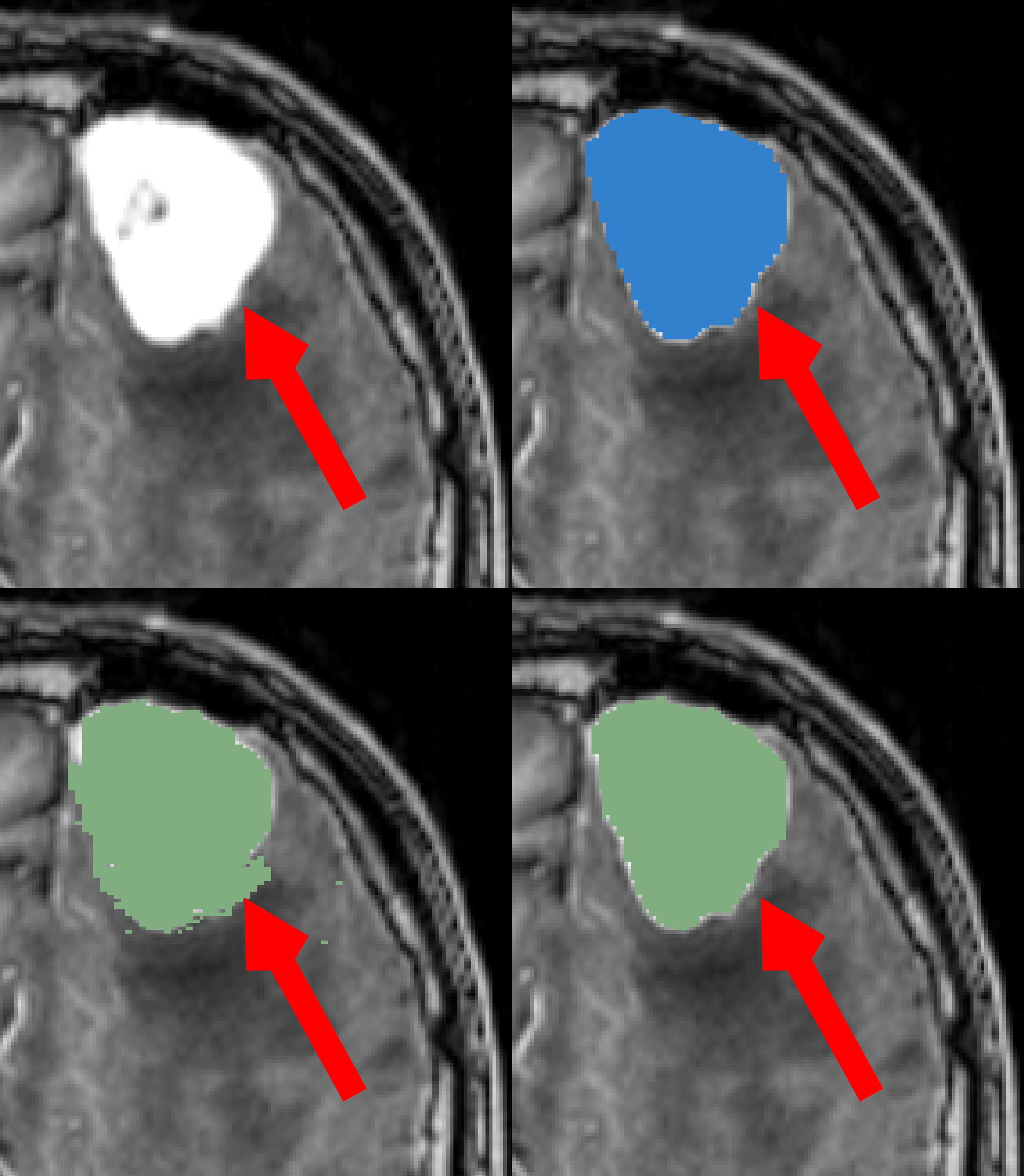}
\caption{Illustration of noise in the ground truth from the use of 3D Slicer, indicated by a red arrow. Original image (top left), prediction with PLS-Cfg4 (top right), ground truth used for training (bottom left), fully manual ground truth from a second expert (bottom right).}
\label{fig:study_GT_quality}
\end{figure}

The ground-truths were originally not created in a pure manual fashion but rather with the assistance of semi-automatic methods from 3D Slicer for time-efficiency purposes. As a result, the presence of some noise in the ground truth has been identified, as illustrated in Fig.~\ref{fig:study_GT_quality}. While such noise is not detrimental since our models appear to be robust and do not generate small artifacts, cleaning the ground truth should lead to a slightly better model and increase the obtained Dice by some percents.

\section{Discussion}
\label{sec:discussion}
The dataset used in this study is larger than any previously described dataset in a meningioma segmentation paper. MRI investigations have been performed using multiple scanners in seven different hospitals, reducing potential biases and preventing overfitting issues. In addition, the smaller meningiomas from the outpatient clinic exhibit a wider range of size and location in the brain which enables our models to be more robust. The identification of slight noise in the ground-truth, due to the use of an external software to facilitate the task, is a slight inconvenience and should be adjusted. Putting the noise aside, the manual annotations from both experts were matching almost perfectly ensuring the overall quality of our dataset.\\
Overall, the PLS-Net architecture provides the best performances with no additional efforts or adjustments. Smarter training schemes are necessary to be implemented for the U-Net architecture, providing a clear speed-up with no impact on the segmentation performances. Nevertheless, due to the slabbing strategy and the lack of global information, reaching the same performances as with PLS-Net seems unachievable. In a local slab a part of a meningioma might appear  similar to other hyperintense structures, making the network struggle. In the future, and with the increasing access to medical data, such training schemes would need to be even more complex with no clear benefit until GPUs are large enough to fit high-resolution MRI volumes in combination with deep architectures.While using a batch-size of 4 is not detrimental for reaching an optimum when training the PLS-Net architecture, batch normalization layers are not optimally put to use which can be one explanation regarding the difference between the validation loss and the actual results. This discrepancy can also originate from not computing the Dice score on exactly the same images. The difference in resolution, spacing, and extent between a preprocessed MRI volume and its original version can amount to Dice score variations when computed using the exact same ground-truth and detection.
Trained models are also robust to the ground-truth noise since predictions do not exhibit the same patterns of small fragmentation. Nevertheless, cleaning the ground-truth is imperative to generate better models since the loss function is based on the Dice score computation.\\
Considering the trade-off between model complexity, memory consumption, and training/inference speed, the PLS-Net architecture is clearly superior for the task of single class segmentation. While meningiomas can be expressed in a large variety of shapes and sizes, their localization in the brain is important. Such information can only be captured by processing the entire MRI volume at once and will be somewhat lost when using a slabbing scheme. In addition, and given that only one class is to be segmented, the huge amount of trainable parameters from U-Net is superfluous. The limitations of the PLS-Net would be apparent if multiple classes were to be segmented.
Compared with the U-Net architecture, the use of the lightweight PLS-Net architecture proves to be better both in terms of segmentation performances but also in terms of training and inference speed. Dividing tenfold the training time is especially relevant with the increase in data collection and the need for models re-training on a regular basis. Different training schemes and data augmentation techniques can also be investigated in a relatively short amount of time.\\
On top of the neural network architecture choice, using mixed precision during training played an essential role to drastically reduce training time. Given the reduced memory footprint, larger-resolution input samples or larger batch size can be investigated.
Having identified that small meningiomas are often missed, increasing the input resolution should help the network finding smaller objects. The downside would be a longer training time, and potentially difficulties to converge if the batch size has to be lowered all the way to 1. Increasing the variability or ratio of small meningiomas in the training set might also steer the network in the correct direction. Lastly, hard-mining could be a potential alternative after careful analysis of the training samples. In any case, using mixed precision by default in the future seems to be a promising strategy in many applications.\\
Compared to previous studies, similar results are obtained using only one MR sequence and without heavy preprocessing (i.e., bias correction, registration to MNI space and skull stripping). In a lightweight framework, and with a shallow multi-scale model, a new patient's MRI can be processed in at most 2 minutes with CPU making it interesting for clinical routine use.\\
In this study, directly benchmarking our models' performances with state-of-the-art results was not possible due to a lack of a publicly available meningioma dataset. Most previous brain tumor segmentation studies have used the BRATS challenge dataset which contains only glioma patients. The few studies focusing on meningioma segmentation used considerably smaller datasets with at most 56 patients in the test set, while not being openly accessible at the same time. Nevertheless, the size of our dataset is on-par with the BRATS challenge dataset which contains 542 patients overall and a fixed test set of 191 patients as of 2018. In addition, we do report our results after performing 5-fold cross-validations, which provides better insight into the model's reproducibility, robustness, and capacity to generalize, compared to a single dataset split into training, testing, and validation sets.

\section{Conclusion}
\label{sec:conclusion}
In this paper, we investigated the task of meningioma segmentation in T1-weighted MRI volumes. We considered two different fully convolutional neural network architectures: U-Net and PLS-Net. The lightweight PLS-Net architecture enables both high segmentation performances while having a very competitive training and processing speed. Using multi-scale architectures and leveraging the whole MRI volume at once impacts mostly the F1-score which is beneficial for automatic diagnosis purposes. Smarter data balancing and training schemes have also shown to be necessary in order to improve performances. In future works, improved multi-scale architectures specifically tailored for such tasks should be explored, but improvements could also come from better data analysis and clustering.

\subsection*{Disclosures}
The authors declare that the research was conducted in the absence of any commercial or financial relationships that could be construed as a potential conflict of interest.\\
Informed consent was obtained from all individual participants included in the study.

\subsection* {Acknowledgments}
This work was funded by the Norwegian National Advisory Unit for Ultrasound and Image-Guided Therapy (usigt.org).

\bibliographystyle{unsrt}  
\bibliography{references}

\end{document}